\documentclass[%
 reprint,
 showkeys,
superscriptaddress,
 amsmath,amssymb,
]{revtex4-1}
\usepackage{graphicx}
\usepackage{dcolumn}
\usepackage{bm}
\usepackage{siunitx}\usepackage{bm}
\usepackage{siunitx}

\usepackage[utf8]{inputenc}
\usepackage{caption}
\usepackage{color,soul}
\begin{document}
\preprint{APS/123-QED}
\title{From hidden metal-insulator transition to Planckian-like dissipation by tuning disorder in a nickelate}
\author{Qikai Guo}
\affiliation{Zernike Institute for Advanced Materials, University of Groningen, The Netherlands}
\author{Beatriz Noheda}
\email{b.noheda@rug.nl}
\affiliation{Zernike Institute for Advanced Materials, University of Groningen, The Netherlands}
\affiliation{CogniGron center, University of Groningen, The Netherlands}
\date{\today}

\begin{abstract}
Heavily oxygen deficient NdNiO$_3$ (NNO) films, which are insulating due to electron localization, contain pristine regions that undergo a hidden metal-insulator transition. Increasing oxygen content increases the connectivity of the metallic regions and the metal-insulator transition is first revealed, upon reaching the percolation threshold, by the presence of hysteresis. Only upon further oxygenation is the global metallic state (with a change in the resistivity slope) eventually achieved. It is shown that sufficient oxygenation leads to linear temperature dependence of resistivity in the metallic state, with a scattering rate directly proportional to temperature. Despite the known difficulties to establish the proportionality constant, the experiments are consistent with a relationship 1/$\tau$= $\alpha k_B T/\hbar$, with $\alpha$ not far from unity. These results could provide experimental support for recent theoretical predictions of disorder in a two-fluid model as a possible origin of Planckian dissipation.

\end{abstract}

\maketitle
\section*{INTRODUCTION}

Among the perovskites, the rare-earth nickelates (RENiO\textsubscript{3}) are interesting because of their tunable metal-insulator transition (MIT)  \cite{canfield1993extraordinary,catalano2018rare,amboage2005high,shi2013correlated, ha2013electrostatic, bubel2015electrochemical,ojha2019anomalous,dominguez2020length}. Recent seminal works directly imaging the evolution of metal and insulating regions, have greatly contributed to a better understanding of the phase transition in these materials\cite{mattoni2016striped,preziosi2018direct, post2018coexisting,li2019scale,lee2019imaging} but important questions remain, such as why are the metallic regions still present at low temperatures \cite{mattoni2016striped} or what is the origin of the critical behaviour observed at the (first-order) phase transition \cite{li2019scale}. RENiO\textsubscript{3} are also attracting renewed attention due to the discovery of a superconducting phase in the family of the infinite-layer nickelates \cite{li2019superconductivity,osada2020superconducting}, which are an oxygen deficient version of the perovskite nickelates, obtained by removal of a full layer of oxygen atoms. However, most often, $V_o$ in nickelates are randomly distributed through the lattice, with a different effect on the transport properties \cite{wang2016oxygen,heo2017influence,onozuka2017reversible,kotiuga2019carrier,zhang2019flexible}. In particular, these randomly distributed $V_o$ are known to give rise to electron localization in defect states \cite{wang2016oxygen,heo2017influence,harisankar2019strain,kotiuga2019carrier}

 In addition, NNO films have been reported to show bad metal behaviour \cite{jaramillo2014origins}, characterized by exceedingly short electron relaxation times; while in  other strongly correlated materials, such as cuprate superconductors or heavy fermions, strange metal behaviour and Planckian dissipation is observed \cite{bruin2013similarity,giraldo2018scale}. This is a recent concept put forward to explain low temperature conductivity that is independent of the strength of the electron interactions and scales linearly with temperature \cite{patel2019theory,zaanen2019planckian}, defying expected quadratic Fermi liquid behaviour. However, in NNO, linear behaviour of the conductivity has only been rarely claimed  \cite{rajeev1991low,blasco1994structural}. In NNO thin films, other scaling exponent ($1\leq n < 2$) have also been reported \cite{mikheev2015tuning, liu2013heterointerface,kobayashi2015pressure, yadav2018influence,phanindra2018terahertz,stemmer2018non}, which are proposed to arise from spin fluctuations, due to the proximity to a quantum critical point, and to vary with strain \cite{liu2013heterointerface,mikheev2015tuning}. In addition, it has also been shown, in nickelates and in other electron correlated materials, that the apparent exponents experimentally obtained, strongly depend on the quenched disorder induced, for example, by the presence of $V_o$ \cite{herranz2008effect,patel2017non,guo2020tunable}. 

These findings emphasize the importance of controlling $V_o$ in RENiO\textsubscript{3}. In this work, we present a study on the effect of the oxygen content in a NdNiO\textsubscript{3} (NNO) film grown on a YAlO\textsubscript{3} (YAO) substrate under highly-compressive strain conditions, which allows us to decouple the effects of strain and disorder. Appropriate analysis of the evolution of the resistivity ($\rho $) versus temperature ($T$) curves with oxygen content offers a clear picture of the microscopic mechanisms at play, supporting the recent idea that Planckian metal behaviour can arise from a two-fluid scenario in a disorder Hubbard model \cite{lee2019emergent,lee2020microscopic,kumar2005singular,herranz2008effect}. 

\section*{RESULTS}

\begin{figure*}
\includegraphics[width=1\textwidth]{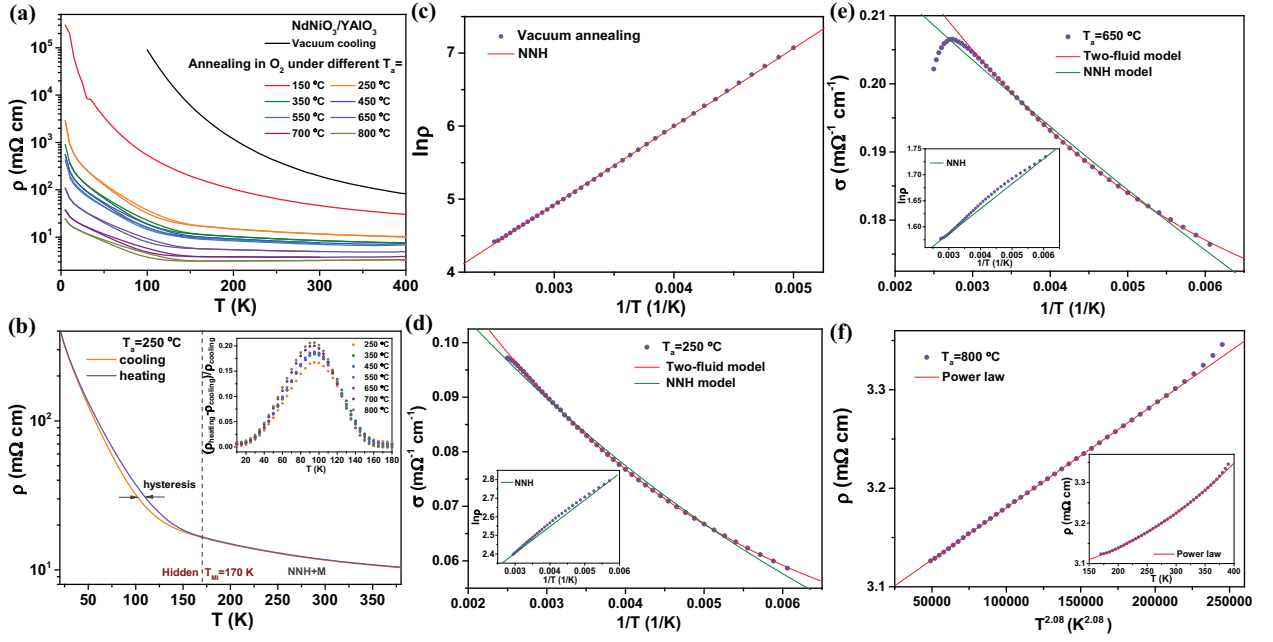}
\caption{\textbf{Modulation of conduction by disorder.} \textbf{a} Temperature ($T$) dependence of resistivity ($\rho$) ($\rho$-$T$ ) during cooling and heating for the same 20 nm thick NNO thin film grown on an YAO substrate, after annealing at increasingly larger temperatures. \textbf{b} $\rho$-$T$ measurements during heating and cooling, after annealing at $T_a$= 250 $^o$C. The inset shows the difference between the heating and cooling curves in \textbf{a}. \textbf{c} Ln($\rho$)-1/$T$ measured in the NNO/YAO film annealed in vacuum, together with the fit to a NNH model for $T$ $\geq$ 170 K. After annealing at $T_a$= 250 $^\circ$C \textbf{d} and 650 $^\circ$C \textbf{e}, the conductivity ($\sigma$) versus $T$ fits best to the two-fluid model of Eq. (\ref{two-fluid model-resistivity})($n$=1). The insets in \textbf{d}-\textbf{e} plotting ln $\rho$-1/$T$, highlight the inadequacy of a NNH fitting model. \textbf{f} The same NNO/YAO film after annealing at 800 $^\circ$C for 1h displays a metal-like $\rho$-$T$  (inset) with a scaling exponent $n$=2.08.} 
\label{fig:resistivity as a function of T}
\end{figure*}

Epitaxial NdNiO\textsubscript{3} (NNO) films are grown on YAlO\textsubscript{3} (YAO) substrates by pulsed laser deposition (see Methods). A 20 nm thick NNO film is subjected to high vacuum conditions right after growth, in order to produce a large amount of $V$\textsubscript{o}. Supplementary Figure 1 shows the atomic force microscopy (AFM) image of the film topography and the X-ray diffraction patterns, attesting the atomically flat, single-crystalline, epitaxial, (001)-oriented nature of the film.
Then, the oxygen content is gradually increased by subjecting the same film to increasingly higher annealing temperatures ($T$\textsubscript{a}) in an oxygen atmosphere. As shown in Supplementary Fig. 1b, the (002)-diffraction peak shows a shift towards increasing angles, corresponding to a reduction in the out-of-plane lattice parameter and manifesting the decrease of $V_o$ content in the film \cite{wang2016oxygen,heo2017influence}.

Figure 1a shows the $\rho$-$T$ curves of the NNO/YAO film measured after different annealing stages, giving rise to various oxygen content levels. The $\rho$-$T$ dependence of the film annealed either in vacuum or in oxygen at a low $T$\textsubscript{a}, show semiconducting behaviour, as expected for a large amount of defects \cite{kotiuga2019carrier}. In order to observe a clear change in the sign of the resistivity derivative, a $T_a$ of 800 $^\circ$C in an oxygen-enriched environment is needed. Under these conditions the film shows a clear metal-to-insulator transition temperature ($T$\textsubscript{MI}) at 170 K (see Supplementary Figure 2). For intermediate states, a progressive decrease of $\rho$ is observed under increasingly $T_a$, in agreement with a systematic reduction in the amount of $V_o$ in the film. However, the temperature dependence of $\rho $ remains insulator-like for $T_a$ up to 700 $^\circ$C.  

In spite of the absence of a change in the temperature derivative of the resistivity, the existence of thermal hysteresis below 170 K, as shown in Fig. 1b and Supplementary Fig. 2a, for $T_a$ above 250 $^\circ$C, is consistent with the coexistence of metallic and insulating regions in an extended temperature range down to 20 K and it is a clear manifestation of the existence of the metal-insulator transition, as expected in pristine NNO \cite{preziosi2018direct}. In the following, we describe how the metallic regions contribute to the resistivity. 

As shown in Fig. 1c and Supplementary Fig. 3 and 4, the $\rho $-$T$ curves of the sample cooled down from the growth in vacuum, as well as after annealing in O\textsubscript{2} at 150 $^\circ$C (with the largest and second largest amount of $V_o$), follow a Nearest Neighbours Hopping (NNH) model for $T$ $\geq$ 170 K.
The NNH conduction model is commonly employed in semiconducting systems \cite{mott1969conduction, shlimak2015hopping} and is described by a simple  activation process: 

\begin{equation}\label{NNH model}
\rho(T)=\rho\textsubscript{0} \exp (E\textsubscript{a}/k\textsubscript{B} T)
\end{equation}
where the $\rho$\textsubscript{0} is a prefactor, $E$\textsubscript{a} is the thermal activation energy of hopping electrons, and $k$\textsubscript{B} is the Boltzmann constant. This model, however, does not apply to the same sample with higher oxygen contents (250 $^\circ$C $ \leq T_a \leq$ 650 $^\circ$C), as seen from the worsening of the NNH fit in Fig. 1d-e and Supplementary Fig. 3b. Under these conditions, the film shows hysteresis in the $\rho$-$T$ curve, as a clear indication of the metal-insulator transition and, thus, of the presence of metallic patches in the samples. The metallic regions are expected to change the resistivity behaviour, which deviates from the NNH model. Indeed, the data are shown to follow a two-fluid NNH+M model \cite{bergman1992physical,kumar2005singular,herranz2008effect} that describes the combined contribution of semiconducting regions, with localized electrons (NNH) and metallic regions, with itinerant electrons (M). The total $\rho$ is then trivially calculated by the parallel combination of these two components as: 

\begin{equation}\label{two-fluid model-resistivity}
\frac{1}{\rho(T)}=\frac{1}{\rho\textsubscript{0}^* \exp(E\textsubscript{a}/k\textsubscript{B} T)}+\frac{1}{\rho^*(0)+A^*T\textsuperscript{n}}
\end{equation}
where $\rho_0^*$ is the prefactor of the NNH contribution, normalized by geometrical factors describing the regions of defective insulating phase, while $A^*$ measures the electron interactions in the metallic regions, normalized by a geometrical factors associated to the pristine metallic phase and $\rho^*(0)$ is the residual resistivity. For convenience, conductivity ($\sigma$) rather than $\rho $ was used in the fitting (see Fig. 1d-e and Supplementary Fig. 4). This parallel arrangement implies that the O-Ni-O bonds in the metallic patches form a connected network in the volume between the electrodes and forming a metallic path through the sample (reaching the percolation threshold). 

 It can also be observed (see inset of Fig. 1b) that both the local $T$\textsubscript{MI} and the thermal hysteresis ($\Delta T$)= 170 K - 20 K remain unchanged during oxygenation, thus, displaying no size dependence as the metallic percolated regions grow in size through the material. This shows that the material of the initial percolating metallic path has the same composition (oxygen content) as the subsequently formed paths, strongly indicating that the metallic regions are made of pristine NdNiO$_3$. Altogether, the data shows that metallic conduction at $T$ $\geq$ $T$\textsubscript{MI} takes place as the stoichiometric regions percolate through the sample, even if the material does not show metallic-like overall behaviour due to the presence of a too large number of $V_o$. The $E_a$ values extracted from the fits are plotted as a function of $T_a$ in Supplementary Fig. 5. We will discuss these values in the next section.

After annealing at 800 $^\circ$C in a oxygen-rich atmosphere, a robust metallic behaviour of $\rho $ is obtained in the formerly oxygen-deficient NNO film (see Fig. 1f). However, the films is not fully oxygenated yet; the $V_o$ content can decrease further by increasing the annealing time, which is demonstrated by the progressive decrease of resistivity shown in Fig. 2a.

As usual, $\rho $- $T$ in the metallic NNO state can be well fit with a power law added to the residual resistivity ($\rho$ - $\rho(0)  \propto AT^n$). The evolution of the extracted values of $n$  (see Supplementary Fig. 6 for the method of extraction) is plotted in Fig. 2b and shows a gradual decrease with decreasing disorder, reaching $n$ = 1 for sufficiently low vacancy content. The results of similar experiments performed in NNO films under tensile strain (on SrTiO\textsubscript{3}) and in quasi-strain free films (on LaAlO\textsubscript{3}) \cite{guo2020tunable} are also plotted in the same figure for comparison.

\begin{figure}
\centering
\includegraphics[width=0.4\textwidth]{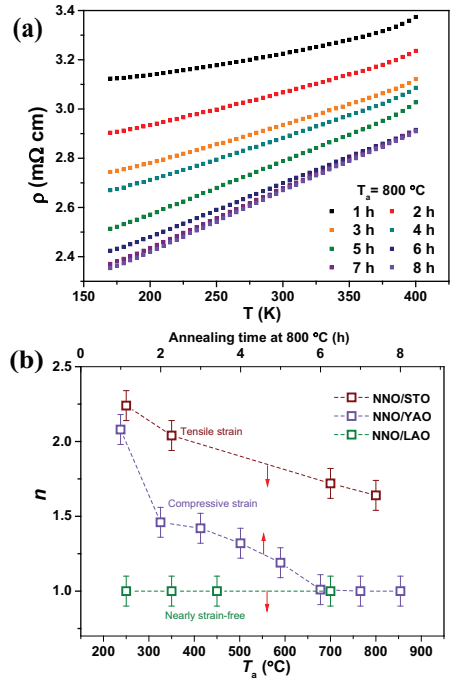}
\caption{\textbf{Tunable resistivity scaling by regulating disorder.} \textbf{a} Resistivity of NNO/YAO film as a function of temperature after annealing in an oxygen-enriched environment at 800 $^\circ$C with various annealing times. \textbf{b} Evolution of exponents $n$ with the annealing process in 20 nm NNO films grown on three different substrates. Data of NNO/LAO and NNO/STO were extracted from previous work \cite{guo2020tunable}.}
\label{fig:metallic resistivity}
\end{figure}

\section*{DISCUSSION}

Our results are consistent with a description of oxygenation as a percolative process (see a sketch in Fig. 3). In the initial state, the NNO film after vacuum cooling possesses a large amount of Ni\textsuperscript{2+}, acting as a semiconducting matrix (grey regions in Fig. 3). The $\rho $-$T$ then follows a NNH model for $T> $170 K (which is the $T$\textsubscript{MI} of the nearly-stoichiometric film), while a Variable Range Hopping (VRH) mechanism is found to dominate in the lower temperature range (see Supplementary Fig. 7), indicating a change in hopping behaviour as a precursor of the metal-insulator transition, even if no metallic behaviour is yet observed. For $T_a$= 250 $^\circ$C, the first percolation paths of stoichiometric regions form across the film. Above $T$\textsubscript{MI}, these paths are metallic but, as they are surrounded by the dominant insulating (defective) matrix, the film still shows semiconductor character (d$\rho(T)$/d$T<$ 0). Indeed, an excellent fit to the data is presented by an NNH+M two-fluid model, which includes itinerant electron metallic resistance in parallel with the resistance contributed by regions that host localized hopping electrons, for all intermediate $T_a$= 250 $^\circ$C - 650 $^\circ$C (see Fig. 1d-e and Supplementary Fig. 4). For each stage of annealing, and upon decreasing the environment temperature, the stoichiometric percolation paths undergo a metal-insulator transition, which is visible as a thermal hysteresis below the local $T$\textsubscript{MI} (marked by green solid lines in Fig. 3). For sufficiently long annealing at 800 $^\circ$C, the metallic regions cover large enough volume ratio to show a sign change in the $\rho$-$T$ slope ((d$\rho(T)$/d$T>$ 0)). This is in agreement with previous reports of NNO films grown on YAO substrates showing a robust metallic $\rho$-$T$ dependence \cite{xiang2013strain,liu2013heterointerface,mikheev2015tuning}.

\begin{figure}
\centering
\includegraphics[width=0.5\textwidth]{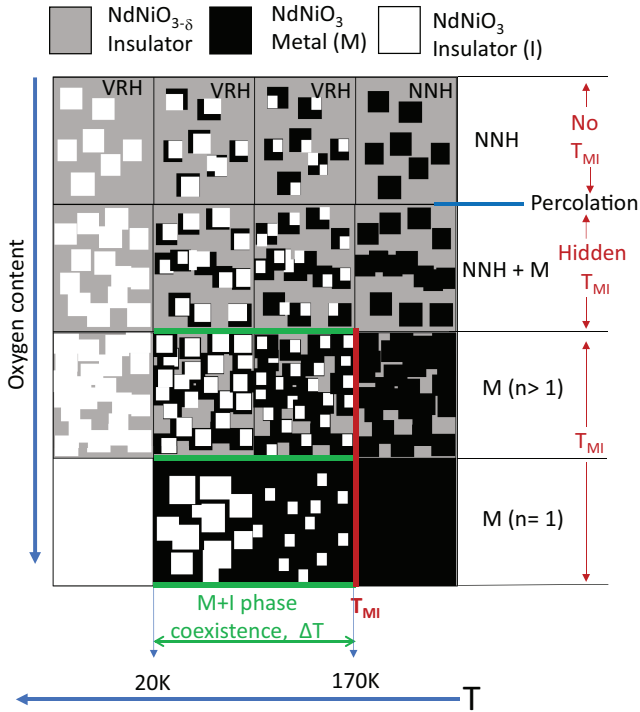}
\caption{\textbf{Illustration of the double percolation model. The sketch shows the evolution of NNO films as a function of oxygen content (vertical) and temperature (horizontal).}  Grey, black and white denote electron-localized (defective) NdNiO\textsubscript{3-$\delta$}, metallic NdNiO\textsubscript{3} and insulating NdNiO\textsubscript{3}, respectively. The evolution of conduction models as a function of oxygen content (from NNH to NNH+M to  M) are denoted. The blue horizontal line signifies the metallic percolation threshold. The green horizontal lines indicate the extent of the hysteresis (coexistence of M and I), while the red vertical line points out the the measurable $T$\textsubscript{MI}.}
\label{fig:schematic of structure}
\end{figure} 

\begin{figure*}
\centering
\includegraphics[width=1.02\textwidth]{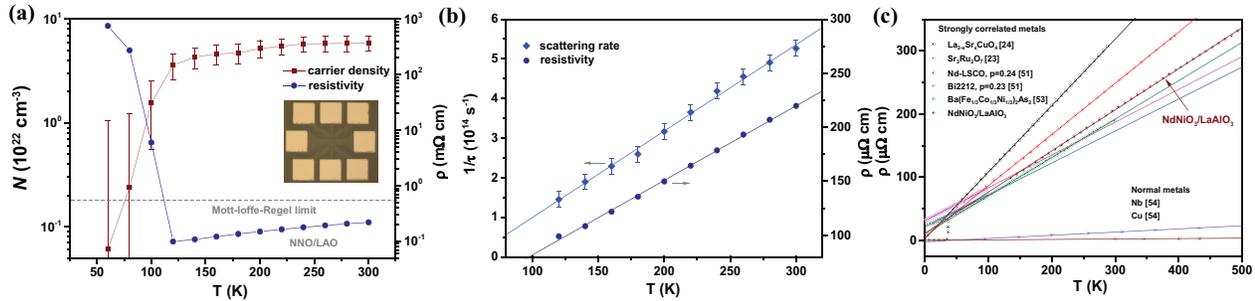}
\caption{\textbf{Strange metal-like $T$-linear resistivity.} \textbf{a} Log of carrier density and log of resistivity of a 5 nm NNO/LAO film as a function of temperature. The inset shows the Hall bar used, with channel dimensions 620 $\times$ 100 $\mu$m\textsuperscript{2}. The dashed line indicates the Mott-Ioffe-Regel limit in nickelates \cite{jaramillo2014origins, mikheev2015tuning}. \textbf{b} 1/$\tau$ as a function of temperature is plotted together with the resistivity and their linear fits.   \textbf{c} The $T$-linear resistivities of reported Planckian metals \cite{bruin2013similarity,giraldo2018scale,legros2019universal,nakajima2020quantum,gunnarsson2003colloquium} together with that of the NNO/LAO film.}
\label{fig:Hall measurement}
\end{figure*}

The progressive oxygenation of the non-stoichiometric matrix is accompanied by a local transformations of different types of polyhedra. In nickelates, oxygen-deficiency leads to a square pyramidal NiO\textsubscript{5} coordination. NiO\textsubscript{4}, with either tetrahedral or square-planar configurations, can be formed if two oxygen atoms of the same octahedra are removed \cite{kotiuga2019carrier}. In heavily deficient cases, the transformation from NiO\textsubscript{4} to NiO\textsubscript{5} coordination could also be expected. This change in the type of polyhedra could be the reason for the observed change in $E_a$ with $T_a$ in the first two states (see Supplementary Fig. 5), different from the constant value of $E_a$ in the intermediate annealing states. $E_a$ $\simeq$ 20 meV is consistent with those reported in the normal semiconducting phases of nickelates \cite{catalan2000metal, medarde1997structural, laffez1999microstructure,granados1992metastable}. However, a clear evolution of the $A^*$-coefficients with decreasing $V_o$ is observed in the intermediate states, indicating a progressive increase of the volume fraction of metallic regions, as expected.

While the metal-insulator transition is clearly of first-order, the order of the  percolation transition under oxygenation still needs investigation. A second-order percolation transition, as in VO$_2$ \cite{liu2016random}, could offer an alternative explanation for the fractal antiferromagnetic structures observed in ref. \cite{li2019scale}. This could perhaps also explain the coexistence of first-order and second-order features also reported in these materials \cite{post2018coexisting}.

Finally, we discuss the temperature scaling of $\rho $ in the recovered metallic state of annealed NNO films on YAO (under compressive strain) in Fig. 2b. In the same figure, we add the NNO films on STO substrates (NNO/STO), under tensile strain, and on LAO substrates (NNO/LAO), nearly strain-free. In the latter, the scaling of $\rho(T)$ is independent of the $T_a$ and gives rise to $n$=1 \cite{guo2020tunable}, as in the bulk material \cite{rajeev1991low,blasco1994structural} and in agreement with ref. \cite{liu2013heterointerface}. We believe this is attributed to a low $V_o$ content compared with the other two systems, which is supported  by values of $\rho $ that are one order of magnitude lower in the metallic state of NNO/LAO film compared with that of the NNO/STO and NNO/YAO films under strain. In the case of tensile strained films (NNO/STO), strain and defect formation are correlated since defects are created during growth as a mechanism to enlarge the lattice and achieve epitaxy \cite{guo2020tunable}. In the current case of compressively strained NNO/YAO, there is no natural tendency for the film to favor vacancies, so these can be created and removed purposely, and the effects of strain and oxygen content can be separated, as shown in Fig. 2b. 

The validity of the $n$ = 1 exponent from above the $T$\textsubscript{MI}= 170 K (and even from $T$\textsubscript{MI} $\simeq$ 100 K in NNO/LAO), significantly smaller than the Debye temperature of the material (420 K \cite{rajeev1991low,liu2013heterointerface}), casts questions about its origin being on electron-phonon interactions and points to the strange metal behaviour of NdNiO$_3$. Strange  metal behaviour is characterized by Planckian dissipation, whose origin is still under debate \cite{patel2019theory}, that supports a resistivity that scales linearly with temperature and is independent of the nature of the electron interactions. Recent theoretical calculations have shown that Planckian metal behaviour can also originate from a disordered Hubbard model (with electron interactions that do not conserve momentum), from which a two-fluid model emerges that consists of localized Sachdev-Ye-Kitaev (SYK) islands interacting with itinerant fermions \cite{lee2019emergent,lee2020microscopic,li2019scale}. Our results seem consistent with this scenario. 

To explicate the underlying physics behind this strange linear-$T$-resistivity,
we use a NNO films (grown under low strain conditions on LAO substrates) with the lowest residual resistivity (10 $\pm$2 $\mu \Omega$ cm) and $\rho$ values well below the Mott-Ioffe-Regel limit \cite{mikheev2015tuning, jaramillo2014origins}, as well as a robust linear-$T$ dependence (from 120 to 500 K). Hall measurements shown in Fig. 4a display a nearly constant value (within error bars) of the carrier density ($N$) in the metallic state, with the expected sharp decrease at and below $T$\textsubscript{MI}. We note that the measured $N$ values are larger than those expected from 1 $e$ per Ni atom. This discrepancy could be due to the existence of both electron-like and hole-like states at the $E$\textsubscript{F} \cite{jaramillo2014origins} or to strong polaronic effects in the metallic state \cite{ha2013hall}, which make reliable measurements of carrier density challenging in nickelates \cite{catalano2018rare}. 

1/$\tau$ values obtained from these measurements, using Drude's formula $\rho=m^*/Ne^2\tau$ and the reported $m^*= 7m_o$ \cite{jaramillo2014origins}, show linear behaviour (see Fig. 4b) and nearly follows the universal dissipation law, 1/$\tau$= $\alpha k_B T/\hbar$, with a constant $\alpha$= 9.3$\pm$ 4.0, larger than that expected in Planckian metals \cite{zaanen2019planckian,bruin2013similarity,legros2019universal,cao2020strange}. However, if a more physically meaningful value of the carrier density is used ($N$= 1.0$\times$ 10\textsuperscript{22}cm\textsuperscript{-3} \cite{jaramillo2014origins,ha2013hall}), a values of $\alpha$= 1.8$\pm$ 0.4, in the proximity of the Planckian dissipation bound ($\alpha$= 1) is obtained. Indeed, the behaviour of NdNiO$_3$ is comparable to that of other reported Planckian metals, in the same temperature range, as shown in Fig. 4c, as well as in Supplementary Table 1 \cite{bruin2013similarity,giraldo2018scale,legros2019universal,nakajima2020quantum,gunnarsson2003colloquium}. Our observations, therefore, suggest that further work is required to fully understand the metallicity of this materials family.

To conclude, in oxygen deficient NdNiO$_3$, metallic regions coexisting with semiconducting defective regions, undergo local metal-insulator transitions, visible above the percolation threshold as thermal hysteresis in the resistivity. Only by further increasing the volume fraction of the pristine state, the metal-insulator transition is unveiled as a change in the slope sign of the resistivity. The $n$= 1 $\rho $-$T$ exponent is recovered upon sufficient oxygenation. This establishes the intrinsic $T$-linear resistivity in metallic NdNiO$_3$ and postulates its strange metal behavior. Although reliable quantification of the carrier density and the value of $\alpha $ in nickelates remains challenging, these results provide experimental support for the disorder-induced two-fluid origin of Planckian dissipation.

\section*{METHODS}

\textbf{Synthesis of oxygen deficient NdNiO\textsubscript{3-$\delta$} film.} An epitaxial NNO film with thickness of 20 nm was deposited on a single-crystal YAlO\textsubscript{3} substrate by pulsed laser ablation of a single-phase target (Toshima Manufacturing Co., Ltd.). Before deposition, the YAO substrates were thermally annealed at 1050 $^\circ$C in a flow of O\textsubscript{2} and etched with DI water to obtain an atomically flat surface with single terminated terraces. The growth was performed at 700 $^\circ$C with an oxygen pressure of 0.2 mbar. The laser fluence on the target was 2 J/cm\textsuperscript{2}. After deposition, the samples were cooled down to room temperature under high vacuum ($\leq 10^{-7}$ mbar) to produce an initial NdNiO\textsubscript{3-$\delta$} state with a large oxygen vacancy content. The thicknesses, crystal orientation and phase purity of the films were assessed using X-ray diffraction by means of $2\theta$ -$\omega$ scans on a Panalytical, Xpert MRD Pro diffractometer.

\textbf{Control of $V_O$ content in film.} The oxygen content in the NdNiO\textsubscript{3-$\delta$} film was tuned by annealing in an oxygen-enriched environment (400 cc/min) with a step-by-step increased temperature. In this annealing process, the $V_O$ in the lattice were gradually refilled with the oxygen atoms from the atmosphere. The content of $V_O$ in the lattice can then decreased gradually by increasing the annealing temperature or annealing time. 

\textbf{Electrical properties measurement.} The electrical transport properties of the film at each stage of annealing were measured between 5 K and 400 K by the van der Pauw method in a Quantum Design Physical Property Measurement System (PPMS), using a Keithley 237 current source and a Agilent 3458A multimeter. For Hall measurement, the films were patterned into Hall bar (channel dimensions 620 $\times$ 100 $\mu$m\textsuperscript{2}) using photolithography  and ion etching. The Pt electrodes with a thickness of 80 nm were fabricated using e-beam evaporation to provide an ohmic contact with the film.

\section*{ACKNOWLEDGEMENTS}

We acknowledge Mart Salverda and Arjan Burema for the support in the construction of devices. We are grateful to Arjun Joshua, Jacob Bass and Henk Bonder for their invaluable technical support. Qikai Guo acknowledges financial support from a China Scholarship Council (CSC) grant and we both acknowledge acknowledge the financial support of the CogniGron research center and the Ubbo Emmius Funds (Univ. of Groningen).

\medskip

\bibliographystyle{jabbrv_apsrev}
\bibliography{Reference.bib}


\section*{Supplementary Material}

\renewcommand{\thefigure}{S1}
\begin{figure*}
\centering
\includegraphics[width=0.9\textwidth]{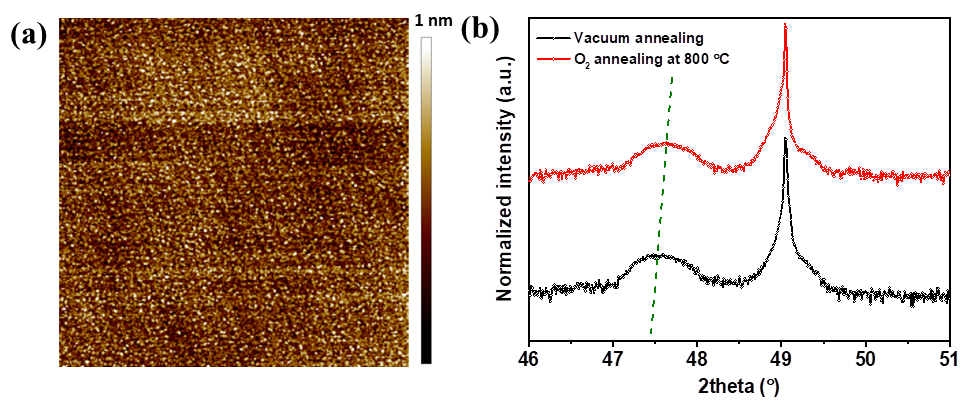}
\caption{(a) AFM topography image of a 3 $\times$ 3 $\mu$m\textsuperscript{2} area in a 20 nm thick NdNiO\textsubscript{3-$\delta$} film grown on YAlO\textsubscript{3} substrate. (b) X-ray diffraction 2$\theta$-$\omega$ scans around the (002) peak of a 20 nm NdNiO\textsubscript{3-$\delta$}/YAO film at different annealing stages. The green dashed line shows the shift of the film peak, indicating an extended out of plane lattice in the film after vacuum annealing due to the larger concentration of oxygen vacancies.}
\label{fig:metallic resistivity}
\end{figure*}

\renewcommand{\thefigure}{S2}
\begin{figure*}
\centering
\includegraphics[width=0.5\textwidth]{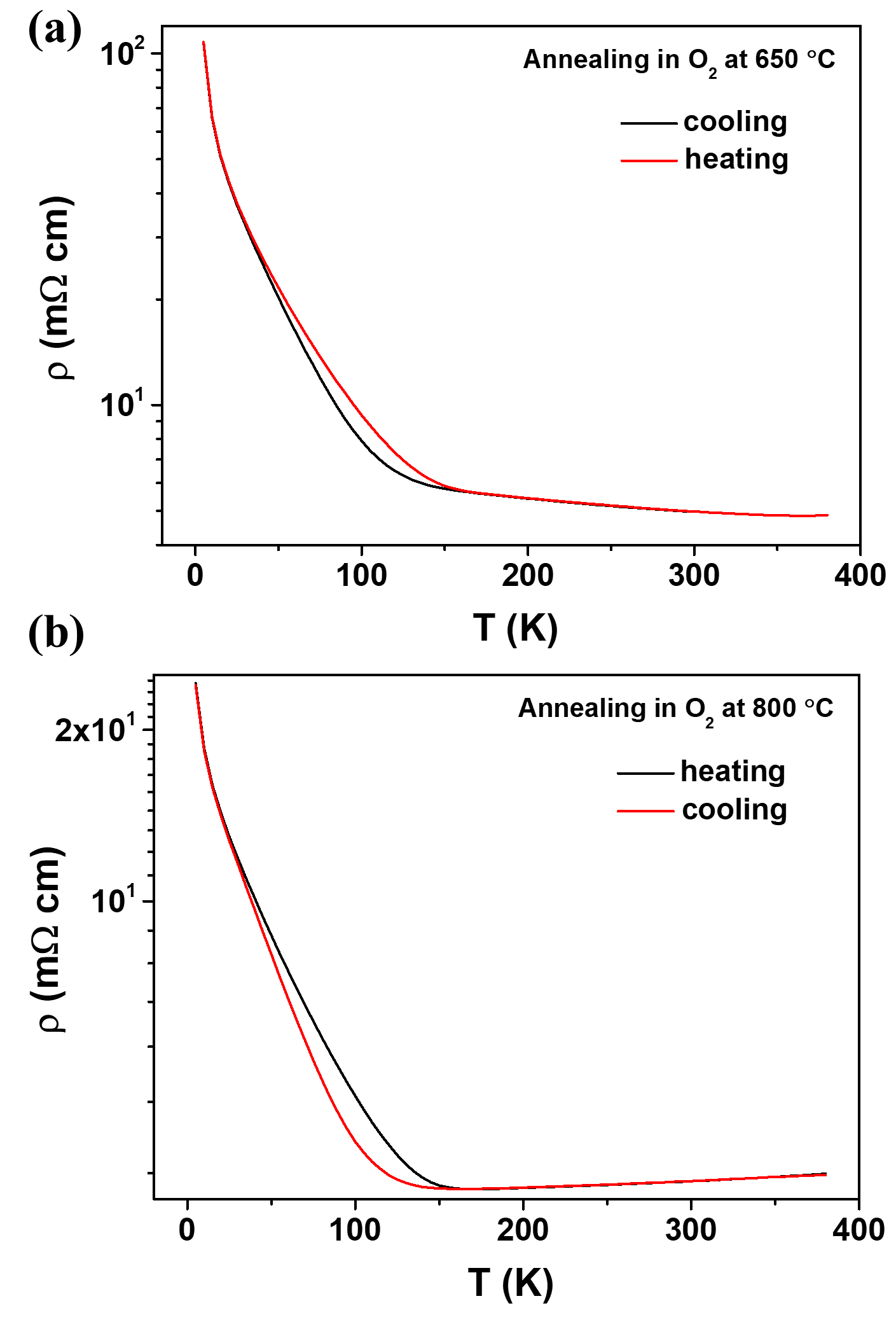}
\caption{Temperature dependence of resistivity for a NdNiO\textsubscript{3-$\delta$}/YAO film after O\textsubscript{2} annealing at (a) 650 $^\circ$C and, subsequently, at (b) 800 $^\circ$C. The resistivity was measured in both the cooling and heating process. A robust metal-insulator transition is recovered in the film after annealing at 800 $^\circ$C, with a $T$\textsubscript{MI}= 170 K. However, the thermal hysteresis is already visible in the film annealed at 650 $^\circ$C when the metallic behavior is still absent.}
\label{fig:metallic resistivity}
\end{figure*} 

\renewcommand{\thefigure}{S3}
\begin{figure*}
\centering
\includegraphics[width=0.9\textwidth]{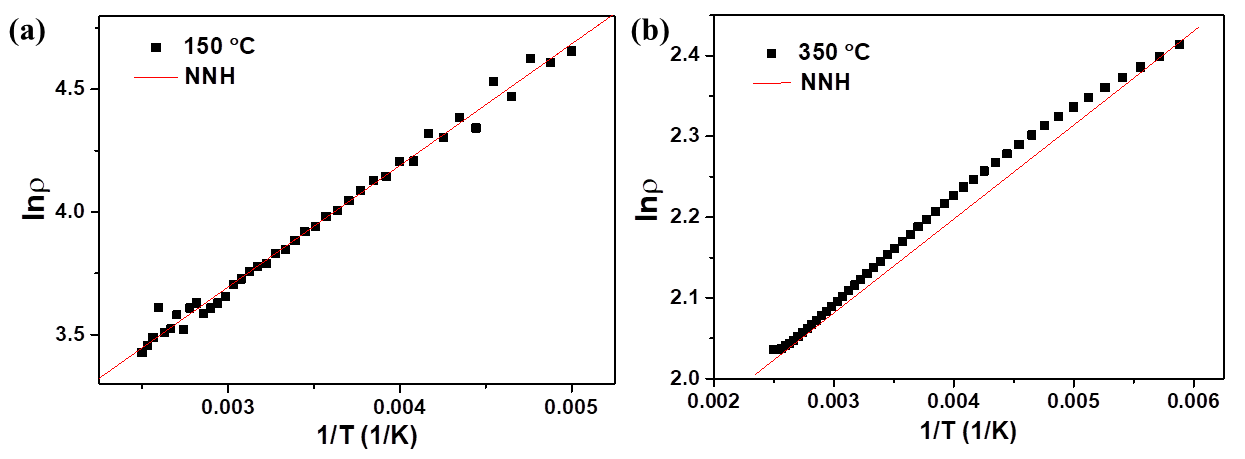}
\caption{$ln\rho$ versus $T^{-1}$ for a NdNiO\textsubscript{3-$\delta$}/YAlO\textsubscript{3} film (a) annealed in 150 $^\circ$C  and (b) 250 $^\circ$C for a temperature range between 180 K and 400 K. The red lines indicate a fit of the experimental data with a Near Neighbours Hopping (NNH) resistivity model.}
\label{fig:metallic resistivity}
\end{figure*}

\renewcommand{\thefigure}{S4}
\begin{figure*}
\centering
\includegraphics[width=0.9\textwidth]{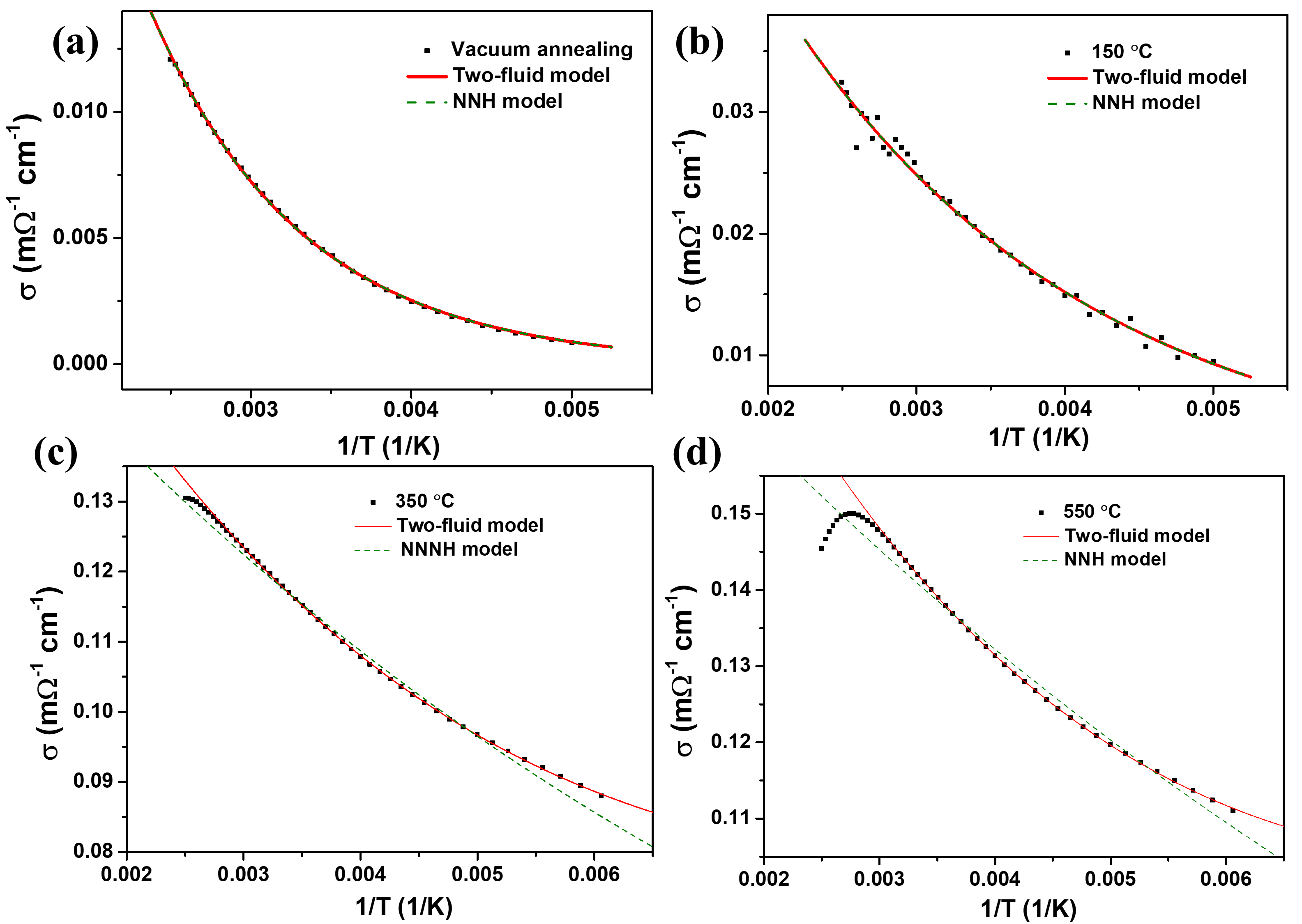}
\caption{$\sigma$ versus $T^{-1}$ plot for  as-prepared NdNiO\textsubscript{3-$\delta$}/YAlO\textsubscript{3} film after vacuum annealing (a) and  O\textsubscript{2} annealing at (b) 150 $^\circ$C, (c) 350 $^\circ$C, and (d) 550 $^\circ$C.  The red solid lines indicate a fit to the experimental data by employing the two-fluid model in Eq.(2), while the green dashed lines are the fit to NNH model. The fits to both models coincide in (a) and (b), indicating the negligible contributions of power law (metallic regions) to the fit. On the contrary, the conductivity in (c) and (d) can only be described by the two-fluid model, emphasizing the importance of  power law (metallic regions) to the fit. }
\label{fig:metallic resistivity}
\end{figure*} 

\renewcommand{\thefigure}{S5}
\begin{figure*}
\centering
\includegraphics[width=0.5\textwidth]{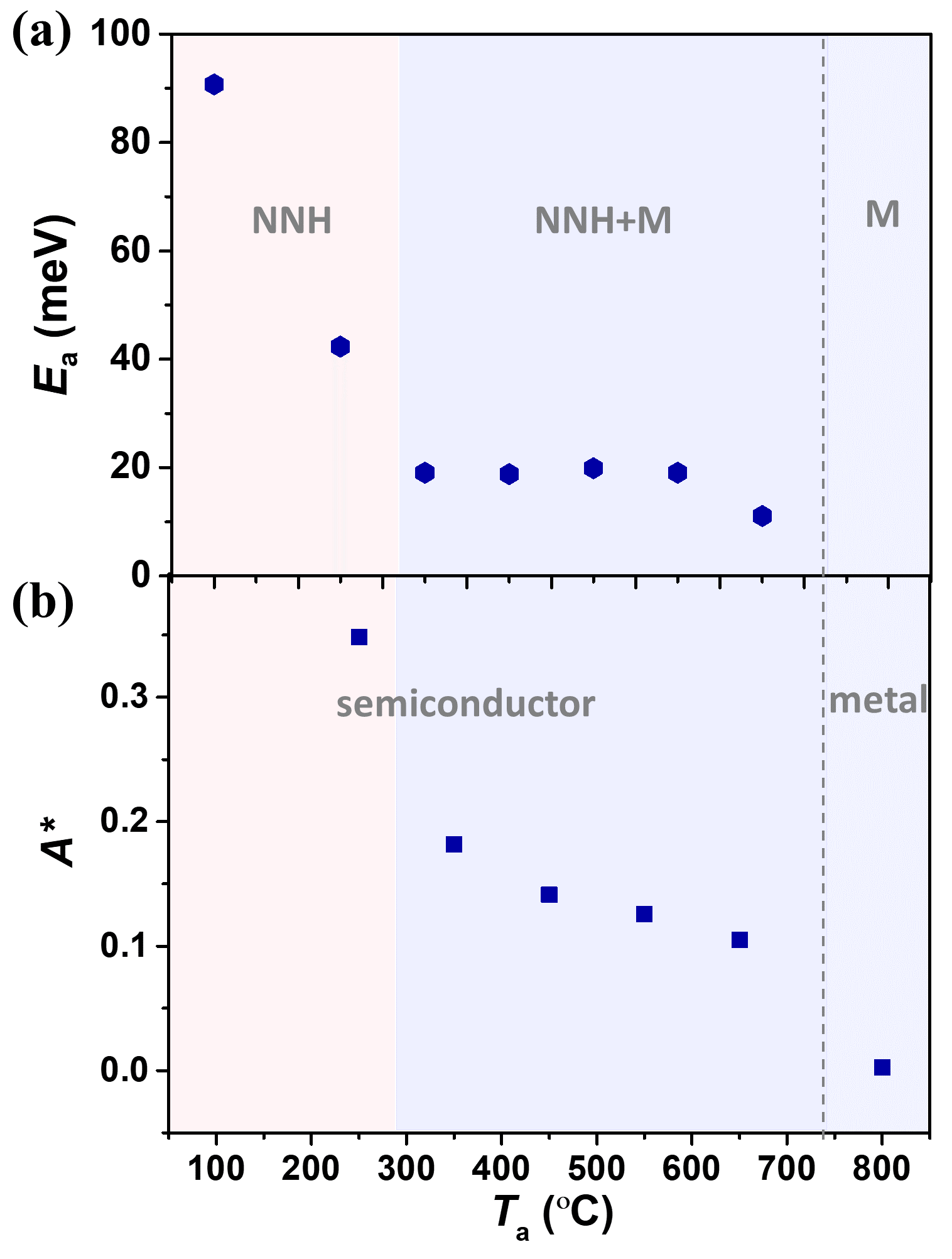}
\caption{Annealing temperature dependence of (a) the thermal activation energy $E_a$ and (b) the $A^*$-coefficient for oxygen deficient NNO/YAO film extracted from the fit to the two-fluid model of Eq. (2). The $A^*$ of 800 $^\circ$C is extracted from the $\rho(T)$ of NNO/YAO film after annealing in oxygen atmosphere for 8 hours, which displays a robust linear $\rho$-$T$ dependence.}
\label{fig:extracted parameters}
\end{figure*}

\renewcommand{\thefigure}{S6}
\begin{figure*}
\centering
\includegraphics[width=0.9\textwidth]{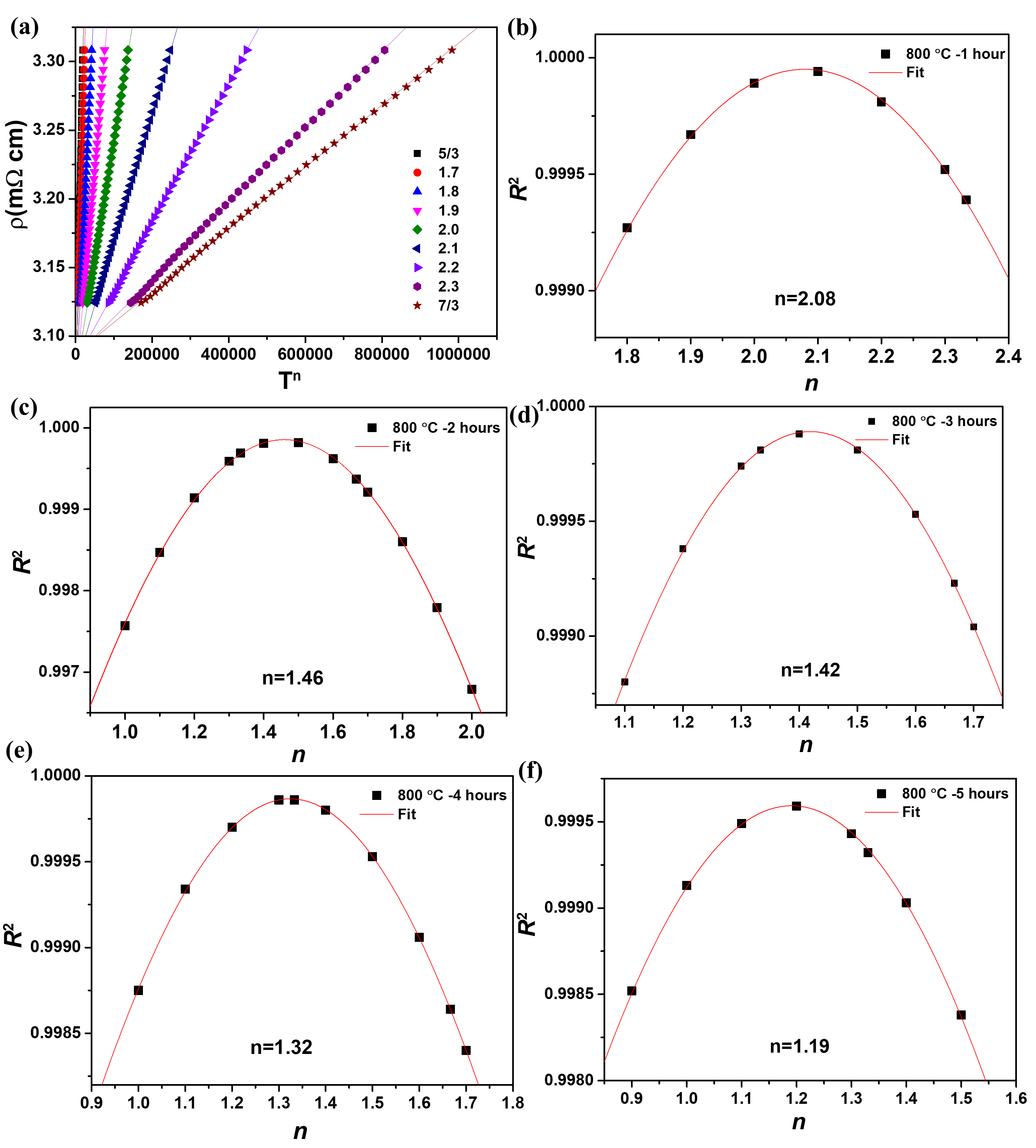}
\caption{Extraction of exponents ($n$). Resistivity of a 20 nm NdNiO\textsubscript{3-$\delta$}/YAlO\textsubscript{3} film (after annealing at 800  $^\circ$C for 1 hour) as a function of $T\textsuperscript{n}$ with different $n$ values.  (b-f) The coefficient of determination ($R$\textsuperscript{2}) as a function of $n$ for the same NdNiO\textsubscript{3-$\delta$}/YAlO\textsubscript{3} film annealed at 800  $^\circ$C for (b) 1 hour, (c) 2 hours, (d) 3 hours, (e) 4 hours, (f) 5 hours, respectively.}
\label{fig:metallic resistivity}
\end{figure*} 

\renewcommand{\thefigure}{S7}
\begin{figure*}
\centering
\includegraphics[width=0.9\textwidth]{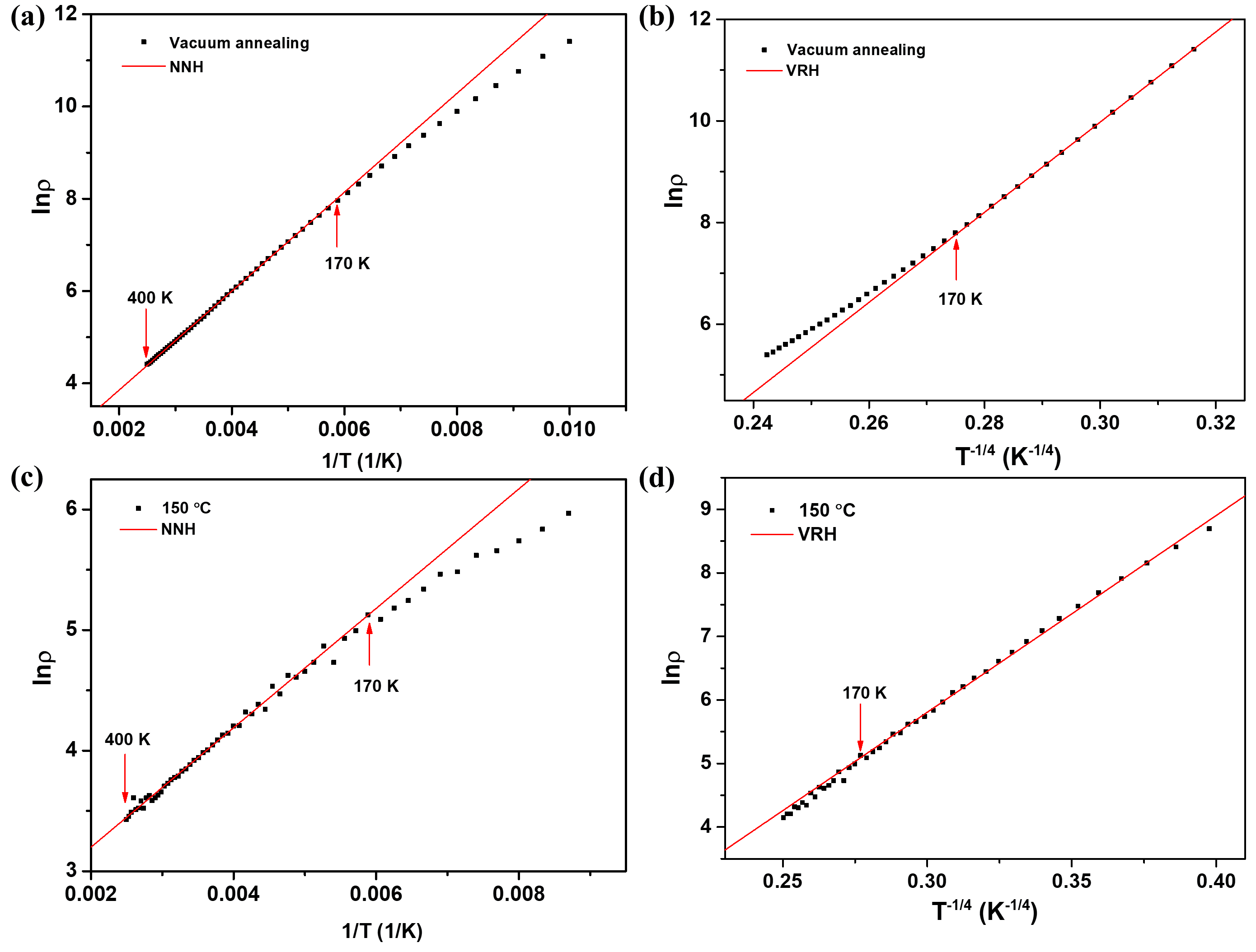}
\caption{(a) $ln\rho$ versus $T^{-1}$ and  (b) $ln\rho$ versus $T^{-1/4}$ for a NNO/YAO film after vacuum annealing. (c) $ln\rho$ versus $T^{-1}$ and  (d) $ln\rho$ versus $T^{-1/4}$ for the same NNO/YAO film after annealed in O\textsubscript{2} at 150 $^\circ$C. The curve is well described by the Near Neighbours Hopping (NNH) model for the temperature range above 170 K, while the Variable Range Hopping (VRH) is applied in the temperature range below 170 K.}
\label{fig:metallic resistivity}
\end{figure*} 

\clearpage

\renewcommand{\thetable}{S1}
\begin{table*}[]
\centering
\caption{{\textbf{Relevant parameters for different reported Planckian metals and the NdNiO$_3$ films presented in this work.} For the definitions of the various parameters, see main text. (Note: the error bar of $N$ is obtained from the uncertainty of the measurement, the error bar of $m^*$ is determined by the different values reported in different works, the error bar of $A$ is obtained from the fit to resistivity, and the error bars of $\alpha$ is calculated from the uncertainty of parameters mentioned above.)}}
\label{tab:my-table}
\resizebox{\columnwidth}{!}{
\begin{tabular}{|l|l|l|l|l|}
\hline
 Material & N(10\textsuperscript{28}m\textsuperscript{-3}) & m\textsuperscript{*}/m\textsubscript{0} & A ($\mu \Omega$ cm/K)  & $\alpha$ \\ \hline
        
Bi2212 (p=0.23)\textsuperscript{[1]} & 0.68 & 8.4 & 0.62 & 1.1\\\hline 
Nd-LSCO (p=0.24)\textsuperscript{[1]} & 0.79 & 12 & 0.49 & 0.7\\\hline 
Sr\textsubscript{3}Ru\textsubscript{2}O\textsubscript{7}\textsuperscript{[2]}] & -- & -- & 1.1 & 1.5\\\hline 
(TMTSF)\textsubscript{2}PF\textsubscript{6}\textsuperscript{[2]} & 0.136 & -- & 0.38 & 0.9\\\hline 
BaFe\textsubscript{2}(P\textsubscript{0.3}As\textsubscript{0.7})\textsubscript{2}\textsuperscript{[68]} & -- & -- & 1.1 & 2.2\\\hline 
CeCoIn\textsubscript{5}\textsuperscript{[2]} & -- & -- & 1.6 & 1.0\\\hline 
Cu\textsuperscript{[2]} & 8.5 & 1.3 & 0.007 & 1.0\\\hline 
Nb\textsuperscript{[2]} & -- & -- & 0.049 & 2.3 \\\hline 
NdNiO\textsubscript{3} & 5.0 $\pm$ 1.0\textsuperscript{[this work]} & 7 $\pm$ 1\textsuperscript{[3]} & 0.6 $\pm$ 0.1\textsuperscript{[this work]} & 9.3 $\pm$ 4.0\\\hline 
NdNiO\textsubscript{3} & 1.0  \textsuperscript{[3,4]} & 7 $\pm$ 1\textsuperscript{[3]} & 0.6 $\pm$ 0.1\textsuperscript{[this work]} & 1.8 $\pm$ 0.4\\\hline%

\end{tabular}
}

\end{table*}

\end{document}